\pdfoutput=1

\documentclass[11pt]{article}

\usepackage[preprint]{acl}

\usepackage{times}
\usepackage{latexsym}

\usepackage[T1]{fontenc}

\usepackage[utf8]{inputenc}

\usepackage{microtype}

\usepackage{inconsolata}

\usepackage{graphicx}

\usepackage{soul}
\sethlcolor{cyan}  

\usepackage[most]{tcolorbox}
\usepackage{enumitem}
\usepackage{booktabs}
\usepackage{balance}
\usepackage{stmaryrd}
\usepackage{pifont}
\usepackage{hhline}
\usepackage{listings}
\usepackage{array}
\usepackage{float}
\usepackage[flushleft]{threeparttable}

\usepackage{mathrsfs}
\usepackage{makecell}
\usepackage{xparse}
\usepackage{wrapfig}
\usepackage{caption}
\usepackage{colortbl}
\usepackage{tabularx} 

\usepackage{epsfig}
\usepackage{multirow}
\usepackage{url}
\usepackage{hyperref}

\usepackage{natbib}
\usepackage{graphicx}

\usepackage{listings}
\usepackage{framed}
\usepackage{xcolor}
\usepackage{color}
\usepackage{geometry}
\usepackage{changepage}

\title{Exposing Out-of-Context Misinformation: A Multi-Agent Approach with Multi-Grained Retrieval}
\title{Detecting Out-of-Context Misinformation via Multi-Agent and Multi-Grained Retrieval}

\title{\paperLogo An Explainable Multi-Agent Framework for Detecting Out-of-Context Misinformation}

\title{\paperLogo An Explainable Agentic Framework for Detecting Out-of-Context Misinformation Leveraging Fine-Grained Retrieval}

\title{A Multi-Agent and Explainable Out-of-Context Misinformation Detection System with Multi-Granularity and Multi-Modal Data Retrieval}

\title{A Multi-Agent Out-of-Context Misinformation Detection System with Multi-Granularity Retrieval}

\title{\textbf{\texttt{EXCLAIM}}: An Explainable Cross-Modal Agentic System for Misinformation Detection with Hierarchical Retrieval}

\author{%
~\textbf{Yin Wu}$^1$,%
~\textbf{Zhengxuan Zhang$^1$},%
~\textbf{Fuling Wang$^1$},%
~\textbf{Yuyu Luo}$^{1,2}$,%
~\textbf{Hui Xiong}$^{1,2}$,%
~\textbf{Nan Tang}$^{1,2}$\thanks{~~Nan Tang and Hui Xiong are the corresponding authors.}
\\
$^{1}$The Hong Kong University of Science and Technology (Guangzhou)\\
$^{2}$The Hong Kong University of Science and Technology
}

\begin{document}
\maketitle
\begin{abstract}
Misinformation continues to pose a significant challenge in today’s information ecosystem, profoundly shaping public perception and behavior. Among its various manifestations, {\bf Out-of-Context (OOC) misinformation} is particularly obscure, as it distorts meaning by pairing authentic images with misleading textual narratives. Existing methods for detecting OOC misinformation predominantly rely on coarse-grained similarity metrics between image-text pairs, which often fail to capture subtle inconsistencies or provide meaningful explainability. While multi-modal large language models (MLLMs) demonstrate remarkable capabilities in visual reasoning and explanation generation, they have not yet demonstrated the capacity to address complex, fine-grained, and cross-modal distinctions necessary for robust OOC detection. To overcome these limitations, we introduce \textbf{\texttt{EXCLAIM}}, a retrieval-based framework designed to leverage external knowledge through multi-granularity index of multi-modal events and entities. Our approach integrates multi-granularity contextual analysis with a multi-agent reasoning architecture to systematically evaluate the consistency and integrity of multi-modal news content. Comprehensive experiments validate the effectiveness and resilience of \textbf{\texttt{EXCLAIM}}, demonstrating its ability to detect OOC misinformation with {\bf 4.3\%} higher accuracy compared to state-of-the-art approaches, while offering explainable and actionable insights.

\end{abstract}
\section{Introduction}

Exponential growth in social media platforms has revolutionized the accessibility, cost-efficiency, and speed of news dissemination through multi-modal channels~\citep{akhtar2023multimodal}. Despite these advancements, the same mechanisms have facilitated the rapid spread of misleading or fabricated information. Among the most concerning forms of misinformation is \textbf{Out-of-Context (OOC)} news~\citep{qi2024sniffer,papadopoulos2024verite}, where authentic images are deliberately misrepresented by associating them with incorrect or deceptive contextual information. For example, during the recent U.S. presidential election, malicious actors exploited this technique by pairing genuine election-related images with unrelated or misleading textual descriptions, constructing false narratives designed to manipulate voter perceptions. Such tactics not only distort public opinion but also erode trust in credible sources of information.

\paragraph{Existing Solutions and Their Limitations.} 
Existing methods for detecting OOC misinformation can be broadly categorized into pre-MLLM solutions and MLLM-based approaches. Pre-MLLM solutions primarily relied on unimodal or multi-modal semantic similarity metrics~\citep{zhou2020similarity, abdelnabi2022open}. These methods focus on extracting semantic features from image-text pairs or simple entity matching but often lacked the ability to analyze context or generate explanations. 

Recently, MLLMs have been used to detect OOC misinformation and generate explanations. However, they either use their learned world knowledge (i.e., in-context learning) or retrieve coarse-grained information (i.e., entire documents)~\citep{mu2023self, qi2024sniffer}. The former is error-prone due to the hallucination of MLLMs, and the latter is hard to precisely retrieve fine-grained information (e.g., a person or an event) required for OOC detection.

\paragraph{Rethinking OOC Detection: Insights from Human Expertise.}
Building on the limitations of existing methods, it becomes crucial to draw inspiration from how human fact-checkers tackle OOC misinformation. Human specialists employ a systematic verification process~\citep{politifact2018,factcheck} that goes beyond surface-level analysis, incorporating multi-granularity reasoning and explainable conclusions. This process includes retrieving information from diverse sources, cross-validating details, and reasoning about timelines, contexts, and inconsistencies. For instance, an expert may trace the origin of the given image, compare it against trusted sources, and evaluate its contextual alignment within a broader narrative. 
This workflow is iterative and hierarchical: individual experts independently analyze evidence, but their findings often converge through peer review to form a consensus. The explanability and adaptability inherent in this process highlight the need for computational frameworks that emulate these characteristics. Existing approaches fail to address this gap, necessitating a rethinking of OOC detection systems.

\paragraph{Our Proposal: An Explainable Multi-Agent Framework for OOC Detection.}
To address these challenges, we propose \textbf{\texttt{EXCLAIM}} (\textbf{EX}plainable \textbf{C}ross-Moda\textbf{L} \textbf{A}gent\textbf{I}c System for \textbf{M}isinformation Detection). Inspired by human detection methods, \textbf{\texttt{EXCLAIM}} introduces a systematic and explainable approach to OOC detection. At its core is a self-constructed database that integrates multi-granularity information across sources and modalities, enabling robust retrieval and context-aware analysis.
\textbf{\texttt{EXCLAIM}} employs a multi-agent architecture that mirrors the systematic reasoning used by human experts. The agents collaboratively retrieve relevant data, analyze multi-modal inconsistencies, and synthesize findings, ensuring both efficiency and explanability. The explanations generated by \textbf{\texttt{EXCLAIM}} are highly aligned with those produced by human experts, providing explainable and trustworthy insights into the detection process.

\paragraph{Contributions.} Our notable contribution can be summarized as follows:
\begin{itemize}
    \item  We propose a {\bf construction method} and introduce a {\bf self-constructed multi-granularity database} for OOC detection, which encapsulates both entity and event-level information from existing news and knowledge.

    \item We propose a {\bf multi-agent} OOC detection framework \textbf{\texttt{EXCLAIM}}, which cross-validates multi-granularity information with input news to be checked. It can not only perform sophisticated reasoning and OOC detection, but also give {\bf explanation} that the news is OOC based on which information source.

    \item Extensive experiments validate the robustness and effectiveness of our framework across various types of OOC misinformation. \textbf{\texttt{EXCLAIM}} achieves a \textbf{4.3\%} improvement in accuracy compared to the SOTA explainable methods, demonstrating its superior performance in detecting OOC misinformation.
\end{itemize}

\section{Related Work}

\subsection{Pre-MLLM Misinformation Detection}
Early misinformation detection research focused on semantic feature extraction from news content, but as fake and real news became semantically indistinguishable, researchers shifted towards leveraging external knowledge~\citep{zhou2020survey}. This transition led to various knowledge-enhanced approaches, such as CompareNet~\citep{hu2021compare}, which constructs directed heterogeneous document graphs to compare news content with knowledge bases through entity extraction. Building on this foundation, recent work has emphasized knowledge retrieval for more precise fact-checking. Notable advances include a retrieval-augmented generation framework for evidence-grounded outputs~\citep{yue2024evidence}, a unified inference framework integrating multiple evidence sources~\citep{Wu2024Unified}, and document-level claim extraction methods~\citep{deng2024document}. While these approaches have demonstrated the value of external knowledge in improving detection accuracy~\citep{dun2021kan, hu2021compare, qian2021knowledge}, they have yet to fully address the utilization and interaction between information at different granularities.

\begin{figure*}[t!]
    \centering
    \includegraphics[width=\linewidth]{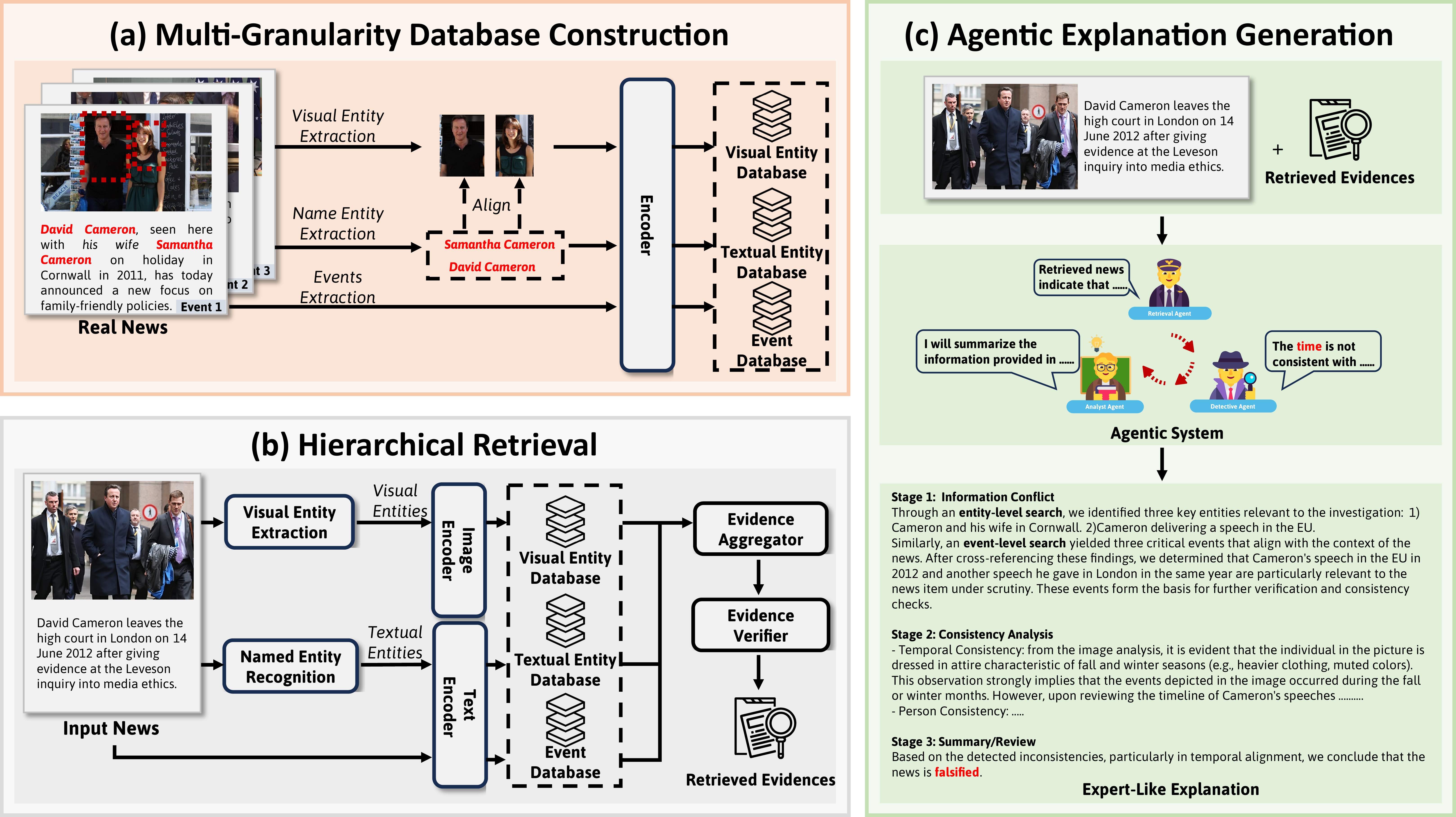}
    \caption{The \textbf{\texttt{EXCLAIM}} Architecture.}
    \label{fig:framework}
\end{figure*}

\subsection{MLLM Assisted Misinformation Detection}

While traditional approaches to misinformation detection have predominantly focused on unimodal data, recent advances in vision-language models have significantly enhanced the ability to detect multi-modal inconsistencies. For instance, \citet{abdelnabi2022open} extended the NewsCLIPpings dataset \citep{luo2021newsclippings} by incorporating external evidence and introduced the Consistency Checking Network (CCN), which evaluates both image-to-image and text-to-text consistency. Similarly, the Stance Extraction Network (SEN) \citep{yuan2023support} builds on the same encoders but improves performance by clustering external evidence semantically to infer its stance towards the claim. SEN also enhances consistency detection by capturing the co-occurrence of named entities across textual and external evidence.

Further advancing the field, the Explainable and Context-Enhanced Network (ECENet) combines coarse- and fine-grained attention mechanisms to model multi-modal feature interactions~\citep{zhang2024escnet}. ECENet utilizes different encoders to jointly process textual and visual entities, offering more nuanced detection of inconsistencies. In addition, 
SNIFFER~\citep{qi2024sniffer} addresses both ``internal consistency'' in image-text pairs and ``external consistency'' with external evidence. 
A parallel line of research has focused on developing interpretable multi-modal architectures for misinformation detection. These approaches~\citet{liu2023interpretable,ma2024interpretable,zhang2023interpretable}. emphasize transparent decision-making processes while maintaining high detection accuracy.

\section{Methodology}

We propose \textbf{\texttt{EXCLAIM}} (\textbf{EX}plainable \textbf{C}ross-Moda\textbf{L} \textbf{A}gent\textbf{I}c System for \textbf{M}isinformation Detection), a multi-granularity framework for OOC detection, integrating both fine-grained entity-level and coarse-grained event-level information. 
As shown in Figure~\ref{fig:framework}, our approach consists of three core components: \textbf{a) Multi-granularity Database Construction,} where visual and textual entities are extracted and aligned using a lightweight MLLM, alongside the storage of event-level information extracted from news captions; \textbf{(b) Hierarchical Retrieval,} which retrieves both entity-level and event-level data through a unified encoding mechanism; and \textbf{(c) Agentic Explanation Generation,} leveraging specialized agents to analyze the consistency of the retrieved evidence and generate explainable OOC detection results.
\subsection{Multi-Granularity Database Construction}
The evidence storage module is designed to extract, align, and store visual and textual entities, as well as event-level information from news items for efficient retrieval using similarity search of Faiss~\citep{douze2024faiss}. Both visual and textual inputs are processed through specialized models, and only aligned entities are stored for rapid querying.

\subsubsection{Multi-Modal Entity Extraction}
\label{ssse: entity extraction}

Given a news item $N = (I, T)$, where $I$ represents the news image and $T$ is the news caption, the system first extracts visual and textual entities. A multi-modal entity is defined as a pair consisting of a visual entity and its corresponding textual entity, where both refer to the same real-world object or concept. Specifically, a multi-modal entity is represented as $(v_i, t_i)$, where $v_i$ is a visual entity extracted from $I$, and $t_i$ is a textual entity extracted from $T$.
Visual entity $v$ is extracted from $I$ using the YOLO v8 Instance segmentation model $M_{\text{YOLO}}$~\citep{yolov8_ultralytics}, producing a set of detected visual entities $V$. Textual entity $t$ is extracted from $T$ using the spaCy NER model $M_{\text{NER}}$~\citep{spacy2}, resulting in a set of textual entities $T$. Thus, the sets of textual entities and textual entities make up the entity set $E=\{(v_1,t_1),....,(v_k,t_k)\}$, where $k$ presents the number of entities.

\subsubsection{Multi-Modal Alignment}
\label{ssse: multimodal alignment}

Before encoding, the system performs multi-modal alignment using a lightweight MLLM. Considering factors such as computational cost, accuracy, and cross-modal understanding capabilities, we selected GPT-4o mini as the model. This model strikes a balance between efficiency and performance, offering robust cross-modal alignment while maintaining low cost. The alignment model $M_{\text{align}}$ gives the similarity between extracted visual entity $v_i$ and textual entity $t_i$, establishing potential mappings between them:

\[
S(v_i, t_i) = M_{\text{align}}(v_i, t_i).
\]

A mapping between a visual entity $v_i$ and a textual entity $t_i$ is considered valid if the similarity score $S(v_i, t_i)$ exceeds a predefined threshold $\tau$. Only entities with valid mappings are retained for further encoding and storage. Entities without sufficient cross-modal similarity are discarded:

\[
E_i = (v_i, t_i) \in E, \quad \text{if} \quad S(v_i, t_i) \geq \tau
.\]

This alignment ensures that only meaningful and relevant visual-textual entity pairs are processed further, reducing storage overhead and improving retrieval precision. 
The mapping information, along with the aligned entities, is saved for future retrieval and analysis.

\subsubsection{Encoding and Storage}

After establishing a valid visual-textual entity $E_j=(v_j,t_j)$, the system proceeds to encode these aligned entities. The visual entities are encoded into high-dimensional feature vectors using the Swin Transformer model $M_{\text{swin}}$~\citep{liu2021swin}, while both the textual entities and the event-level information are encoded using the RoBERTa model $M_{\text{RoBERTa}}$~\citep{liu2019roberta}:
\begin{align*}
Z_V & = M_{\text{swin}}(v_j), \\
Z_T & = M_{\text{RoBERTa}}(t_j), \\
Z_{\text{event}} & = M_{\text{RoBERTa}}(T).
\end{align*}
The encoded representations of the aligned visual entities $Z_V$, textual entities $Z_N$, and event-level information $Z_{event}$ are stored in separate Faiss indices, referred to as $Index_V$, $Index_T$, $Index_{event}$, to enable efficient retrieval.

\begin{figure*}[t!]
    \centering
    \includegraphics[width=\linewidth]{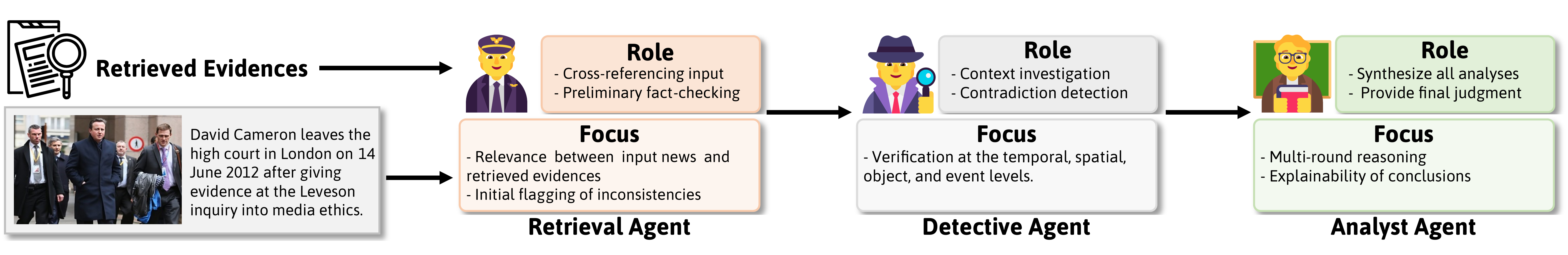}
    \caption{Multi-Agent Detection Workflow. The system employs three agents—Retrieval, Detective, and Analyst—in a sequential pipeline, progressively refining the detection process.}
    \label{fig:multiagents}
\end{figure*}

\subsection{Hierarchical Retrieval}
\label{sse:evidence retrieval}

The Evidence Retrieval module is responsible for retrieving relevant entities, and event-level information from the pre-constructed Faiss index files. This module ensures efficient multi-modal retrieval to support the OOC detection process. The retrieval process consists of two main components: data encoding and retrieval, followed by evidence aggregation and verification.

\subsubsection{Evidence Retrieval}

Given an input news item \( N_{\text{input}} = (I_{\text{input}}, T_{\text{input}}) \), where \( I_{\text{input}} \) represents the image and \( T_{\text{input}} \) is the accompanying caption, the system first performs entity extraction and encoding following the methods described in Sections~\ref{ssse: entity extraction} and~\ref{ssse: multimodal alignment}. Specifically, this process results in the generation of encoded query vectors: \( \mathbf{v}_{\text{query}} \) for the visual component, \( \mathbf{t}_{\text{query}} \) for the textual component, and \( \mathbf{e}_{\text{query}} \) for the event-level information.

Subsequently, the system retrieves the most relevant entities from the respective Meta Faiss indices by calculating the Euclidean distance between the encoded query vectors and the indexed entities. For each modality, the top two nearest entities (in terms of Euclidean distance) are retrieved. This process, referred to as \textbf{top-$k$ retrieval}, is implemented as follows:
\[
\begin{aligned}
    \mathcal{V}_r &= \text{top-$k$}(\mathbf{v}_{\text{query}}, Index_V, k=2), \\
    \mathcal{T}_r &= \text{top-$k$}(\mathbf{t}_{\text{query}}, Index_T, k=2), \\
    \mathcal{E}_r &= \text{top-$k$}(T_{\text{input}}, Index_{event}, k=2).
\end{aligned}
\]
Here, \( \text{top-$k$} \) refers to the process of retrieving the top $k$ entities from the corresponding index \( Index \), ranked by their similarity to the query vector. In this case, we set $k=2$ to retrieve the two most relevant entities. The choice of $k=2$ is motivated by the need to provide diverse yet concise entity representations for downstream tasks.
\subsubsection{Evidence Aggregation and Verification}

After retrieving the relevant visual entities \( \mathcal{V}_r \), textual entities \( \mathcal{T}_r \), and event information \( \mathcal{E}_r \), these are combined by the Evidence Aggregator into a unified evidence set:

\[
\mathcal{E}_{\text{agg}} = \{\mathcal{V}_r, \mathcal{T}_r, \mathcal{E}_r\}.
\]

The aggregated evidence \( \mathcal{E}_{\text{agg}} \) is then passed to the Evidence Verifier, which assesses its consistency and relevance for the OOC detection task. The verifier ensures that there are no duplicates in the retrieved evidence and that the evidence is correctly formatted. After the verification process, \( \mathcal{E}_{\text{agg}} \) is cleaned and validated, ensuring it contains only unique and properly formatted items, ready for further processing.

In summary, this process efficiently encodes and retrieves multi-modal information through Faiss indices, enabling fine-grained entity-level retrieval and broader event-level context for OOC detection.

\subsection{Agentic Explanation Generation}

The Multi-Agent Detection Module forms the core of our OOC detection framework, employing a multi-stage process inspired by Chain-of-Thought (CoT) reasoning. In this framework, each agent is responsible for a distinct phase of the detection pipeline, with the output of one agent seamlessly feeding into the next. This enables not only a sequential but also a highly collaborative workflow, where agents complement and build upon each other's efforts. This structure closely mirrors human reasoning by breaking down complex tasks into smaller, more manageable components, allowing for a more robust and explainable  detection process.


Figure~\ref{fig:multiagents} outlines the roles of the three key agents in our framework: the Retrieval Agent, Detective Agent, and Analyst Agent. These agents operate sequentially to refine the detection process. The Retrieval Agent initiates fact-checking by cross-referencing input news with retrieved evidence, flagging any inconsistencies. The Detective Agent then conducts a deeper investigation, verifying key elements such as time, place, and objects to detect contradictions. Finally, the Analyst Agent synthesizes the previous stages' findings, providing a coherent and explainable conclusion. Through this multi-agent collaboration, \textbf{\texttt{EXCLAIM}} not only achieves high accuracy in detecting out-of-context misinformation but also ensures that the reasoning behind each decision is transparent and explainable. This layered, cooperative approach significantly enhances the robustness and reliability of the overall system.

\subsubsection{Retrieval Agent}

The Retrieval Agent initiates the CoT-inspired process by cross-referencing input news $N_{\text{input}}$ with retrieved evidence $\mathcal{E}_{\text{a}}$. It performs the first consistency check, ensuring alignment between visual and textual entities at both the entity and event levels. Leveraging MLLM’s pre-trained knowledge, the agent identifies significant misalignments, passing flagged inconsistencies as input to the next agent for deeper analysis.

\subsubsection{Detective Agent}

Building on the Retrieval Agent’s results, the Detective Agent conducts a more detailed investigation. It systematically evaluates key elements—\textit{time}, \textit{place}, \textit{person}, \textit{event}, and \textit{object}—to detect contradictions between the retrieved evidence and the input news. For example, it checks if clothing matches the season described or if objects align with the event. This agent’s refined analysis, aligned with CoT reasoning, narrows the scope of potential inconsistencies. The resulting findings are passed to the final agent.

\subsubsection{Analyst Agent and System Output}

The Analyst Agent synthesizes the outputs from the Retrieval and Detective Agents, integrating their findings into a coherent OOC detection report. Acting as an expert reviewer, it provides a well-supported, explainable conclusion, drawing on the cumulative reasoning of prior stages.
The final output of the Analyst Agent is represented as:

\[
O_{\text{final}} = \left( C_{\text{OOC}}, T_{\text{exp}} \right),
\]

\noindent where \(C_{\text{OOC}} \in \{ 0, 1 \}\) indicates the binary classification, with \(C_{\text{OOC}} = 1\) signifying that the news is OOC, and \(C_{\text{OOC}} = 0\) denoting that the news is consistent with the retrieved evidence.
\(T_{\text{exp}}\) provides a comprehensive explanation based on the inconsistencies and contradictions identified during the detection process. This module can facilitate structured, multi-turn dialogue by passing outputs between agents, breaking down OOC detection tasks into manageable steps for robust and explainable outcomes.
\section{Experiments}

\subsection{Experimental Setup}
\subsubsection{Datasets}
We leverage the NewsCLIPpings benchmark~\citep{luo2021newsclippings}, the largest dataset for detecting out-of-context misinformation. This dataset is sourced from the VisualNews dataset~\citep{liu2021visual}, which was initially created for news image captioning. NewsCLIPpings contains news articles from four major outlets: The Guardian, BBC, Washington Post, and USA Today. The dataset is evenly balanced with respect to labels. Its high-quality and diverse sources make it well-suited for large-scale retrieval tasks, ensuring both linguistic richness and broad topic coverage. The dataset is evenly balanced with respect to labels.

Following prior work~\citep{qi2024sniffer}, we report results on the Merged/Balance subset, which ensures an equal distribution of retrieval strategies and positive/negative samples. Specifically, the retrieval strategies are categorized into four types: \textit{Text-Image}, \textit{Text-Text}, \textit{Person Matching}, and \textit{Scene Matching}. This subset includes 71,072 samples for training, 7,024 for validation, and 7,264 for testing. Consistent with~\citep{luo2021newsclippings}, we evaluate performance using accuracy across all samples (All) and separately for the Falsified (Out-of-Context) and Pristine (Not Out-of-Context) samples as evaluation metrics.

\subsubsection{Implementation Details}

\textbf{\texttt{EXCLAIM}} relies on a proprietary multi-granularity database, constructed specifically from the training subset of the NewsCLIPpings dataset. This database is built offline and comprises \textbf{18,305} unique entities and \textbf{71,072} event instances, ensuring comprehensive coverage of the training data. By pre-computing and indexing this data, we enable more efficient retrieval during inference.

To optimize retrieval efficiency, we employ a Faiss index, enabling rapid and scalable access to the multi-granularity data during the reasoning process.Each agent in the multi-agent system is instantiated using GPT-4o, with temperature set to 0.6, ensuring a balance between creativity and consistency across tasks. The model processes inputs with a maximum length of 4096 tokens, allowing it to handle complex reasoning and multi-hop retrieval effectively. This allows us to dynamically generate specialized outputs for entity recognition, event verification, and cross-modal consistency checking. 

\subsubsection{Baselines}
To thoroughly evaluate \textbf{\texttt{EXCLAIM}}’s performance, we compare it to a broad range of SOTA multi-modal models. \textbf{EANN}~\citep{wang2018eann} uses adversarial training to learn event-invariant features, making it robust across various detection scenarios. \textbf{VisualBERT}~\citep{li2019visualbert} processes image-text pairs through a unified transformer, optimizing key tasks such as image-text alignment. \textbf{SAFE}~\citep{zhou2020similarity} enhances prediction accuracy by transforming images into descriptive sentences and applying sentence similarity as an auxiliary loss. \textbf{CLIP}~\citep{radford2021learning} employs separate encoders for images and text, aligned through contrastive learning to ensure semantically related pairs are closely represented. \textbf{CCN}~\citep{abdelnabi2022open} builds on CLIP by incorporating cross-modal consistency checks and external evidence retrieval for improved decision-making. \textbf{DT-Transformer}~\citep{papadopoulos2023DTTransformer} further extends CLIP by introducing additional transformer layers to refine multi-modal interactions, capturing more complex relationships.
\textbf{Neu-Sym Detector}~\citep{zhang2023Neu-Symdetector} combines neural-symbolic reasoning by decomposing text into fact queries and aggregating outputs through a pre-trained multi-modal model. To demonstrate that \textbf{\texttt{EXCLAIM}}'s performance is not solely attributed to the underlying GPT-4o capabilities, we include \textbf{GPT-4o-Latest} in both zero-shot and few-shot settings as strong baselines. These variants represent the direct application of GPT-4o's multi-modal capabilities without the specialized framework components present in EXCLAIM. Finally, \textbf{SNIFFER}~\citep{qi2024sniffer} selects the InstructBLIP as the base MLLM and enhances OOC detection with a two-stage instruction tuning process based on , integrating GPT-4-generated OOC-specific data and external evidence retrieval to improve consistency checks and overall explainability.

\subsection{Main Results}

\begin{table}[t!]
\centering
\caption{Accuracy comparison (\%). The best results for each column are highlighted in bold.}
\label{table1: results}
\begin{tabular}{lccc}
\toprule
\textbf{Method} & \textbf{All} & \textbf{Falsified} & \textbf{Pristine} \\
\midrule
EANN            & 58.1         & 61.8          & 56.2          \\
VisualBERT      & 58.6         & 38.9          & 78.4          \\
SAFE            & 52.8         & 54.8          & 52.0          \\
CLIP            & 66.0         & 64.3          & 67.7          \\
CCN             & 84.7         & 84.8          & 84.5          \\
DT-Transformer  & 77.1         & 78.6          & 75.6          \\
Neu-Sym detector & 68.2        & -             & -             \\
GPT-4o (zero-shot) &         73.8       &   75.5            &      73.4        \\
GPT-4o (few-shot) &          79.2    &     81.1      &    77.4     \\
SNIFFER  & 88.4 & 86.9     & 91.8  \\
\textbf{\texttt{EXCLAIM}} \textbf{(ours)} & \cellcolor{lightgray}\textbf{92.7}  & \textbf{93.3}    & \cellcolor{lightgray}\textbf{92.1}   \\
\hline
\end{tabular}
\end{table}

Experimental results demonstrate \textbf{\texttt{EXCLAIM}}'s superior performance across all evaluation metrics compared to existing approaches. While traditional models trained from scratch (EANN: 58.1\%, SAFE: 52.8\%) and established multi-modal frameworks (CLIP: 66.0\%, VisualBERT: 58.6\%) show limited effectiveness, more recent architectures achieve notable improvements through enhanced mechanisms. CCN (84.7\%) and DT-Transformer (77.1\%) leverage CLIP's foundation with additional consistency checks, while SNIFFER establishes a strong benchmark (88.4\%) through its specialized detection approach. Notably, despite GPT-4o's powerful foundation and advanced reasoning capabilities, its performance peaks at 79.2\% with few-shot learning—a significant improvement over its zero-shot variant (73.8\%) but still substantially below \textbf{\texttt{EXCLAIM}}'s performance, highlighting the limitations of general-purpose language models for specialized detection tasks.

\textbf{\texttt{EXCLAIM}} substantially advances the state-of-the-art with an accuracy of 92.7\%, surpassing SNIFFER by 4.3\% and GPT-4o (few-shot) by 13.5\%. This marked improvement persists across both falsified (93.3\%) and pristine (92.1\%) categories, validating the effectiveness of our multi-agent reasoning framework and multi-granularity database architecture. The significant performance gap between \textbf{\texttt{EXCLAIM}} and these strong baselines, particularly the substantial margin over GPT-4o, underscores the necessity and effectiveness of our specialized architectural design in addressing the unique challenges of OOC detection.



\subsection{Ablation Studies}
\begin{table*}[t!]
\centering
\caption{Ablation Studies on Each Component of \textbf{\texttt{EXCLAIM}} Framework.}
\label{tab:ablation}
\resizebox{\textwidth}{!}{  
\begin{tabular}{ccccc|ccc}
\toprule
\textbf{Analyst Agent} & \textbf{Detective Agent} & \textbf{Retrieval Agent} & \textbf{Event-Level Evidence} & \textbf{Entity-Level Evidence} &  \textbf{All} & \textbf{Falsified} & \textbf{Pristine} \\
\midrule
\textcolor{green}{\ding{51}} & \textcolor{red}{\ding{55}} & \textcolor{red}{\ding{55}} & \textcolor{red}{\ding{55}}& \textcolor{red}{\ding{55}} & 83.6 & 86.3 & 80.9 \\
\textcolor{green}{\ding{51}} & \textcolor{red}{\ding{55}} & \textcolor{red}{\ding{55}} & \textcolor{green}{\ding{51}}& \textcolor{green}{\ding{51}} & 82.7 & 93.1 & 72.3 \\
\textcolor{green}{\ding{51}} & \textcolor{green}{\ding{51}} & \textcolor{red}{\ding{55}} & \textcolor{green}{\ding{51}}& \textcolor{green}{\ding{51}} & 89.2 & 87.5 & 90.9 \\
\textcolor{green}{\ding{51}} & \textcolor{red}{\textcolor{red}{\ding{55}}} &  \textcolor{green}{\ding{51}} & \textcolor{green}{\ding{51}}& \textcolor{green}{\ding{51}} & 88.6 & 91.0 & 86.2 \\
\textcolor{green}{\ding{51}}& \textcolor{green}{\ding{51}}& \textcolor{green}{\ding{51}} & \textcolor{green}{\ding{51}}& \textcolor{red}{\ding{55}} & \textbf{89.2} & \textbf{90.1} & \textbf{88.3} \\
\textcolor{green}{\ding{51}}& \textcolor{green}{\ding{51}}& \textcolor{green}{\ding{51}} & \textcolor{green}{\ding{51}}& \textcolor{green}{\ding{51}} & \cellcolor{lightgray}\textbf{92.7} & \cellcolor{lightgray}\textbf{93.3} & \cellcolor{lightgray}\textbf{92.1} \\
\hline
\end{tabular}
}
\end{table*}

To assess the contributions of each component in \textbf{\texttt{EXCLAIM}}, we conducted ablation experiments (Table~\ref{tab:ablation}). When the \textbf{Retrieval Agent} was absent, relevant evidence was directly provided to the \textbf{Analyst} or \textbf{Detective Agent}, maintaining access to multi-granularity information while bypassing retrieval. This setup allowed us to isolate the impact of each module.

Starting with only the \textbf{Analyst Agent}, which performs high-level reasoning over multi-modal inputs, the system achieved \textbf{83.6\%} accuracy. While effective at detecting falsified content (\textbf{86.3\%}), it struggled with pristine samples (\textbf{80.9\%}). Incorporating the \textbf{Detective Agent}, responsible for fine-grained entity and image analysis, improved falsified content recall to \textbf{93.1\%}, but pristine accuracy dropped to \textbf{72.3\%}, indicating an imbalance when relying solely on entity-level analysis.

The \textbf{Retrieval Agent} played a crucial role in improving performance. Its inclusion boosted overall accuracy to \textbf{89.2\%}, significantly enhancing pristine content detection (\textbf{90.9\%}). Further, integrating \textbf{event-level and entity-level evidence retrieval} maximized performance, with the full \textbf{\texttt{EXCLAIM}} system achieving \textbf{92.7\%} accuracy (\textbf{93.3\%} falsified, \textbf{92.1\%} pristine). These results confirm that all components are essential for optimal \textit{out-of-context} detection, ensuring both accuracy and balance across falsified and pristine samples.

\subsection{Case Study}

\renewcommand\arraystretch{1}
\begin{table*}[t]
\centering
\small
\caption{A Case study comparing GPT-4o and \textbf{\texttt{EXCLAIM}}'s verification capabilities across character. Ground truth labels (\textcolor{cyan}{GT}) are provided.}
\begin{tabular}{lll}
\toprule
\multicolumn{2}{p{26pc}}{{\bf Caption}: People cheer and take pictures as the pope arrives to meet with representatives of the World of Work organization in Ciudad Juarez. \color{cyan}{[GT: Falsified]}} & \makecell[c]{\multirowcell{14}{\includegraphics[width=8pc]{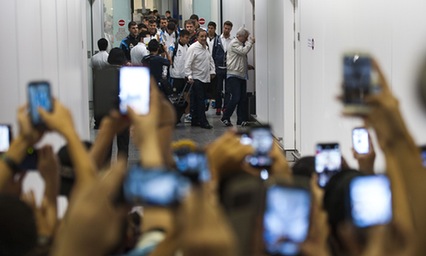}}}  \\
\cmidrule{1-2}
\multicolumn{1}{p{13pc}|}{{\bf GPT-4o}:  The caption describes people cheering and taking pictures as the pope arrives for a meeting with representatives of the World of Work organization in Ciudad Juarez. However, there is nothing in the image that definitively identifies the individual or the specific event being described. Without clear indicators of the pope's presence or explicit evidence linking the scene to Ciudad Juarez or the World of Work meeting, it is \textit{difficult to confirm} whether the caption accurately matches the image.}  
&  
\multicolumn{1}{p{13pc}}{{\textbf{\texttt{EXCLAIM}}}: The image shows a crowd of people holding up phones, seemingly taking pictures or videos. A group of individuals is walking down a hallway, but no specific identifying features are visible. Based on the retrieved news caption, ``The pope smiles as members of the World of Work organization wave to him in Ciudad Juarez,'' we can confirm the \textbf{geographic context} of the described event. However, the \textbf{individuals and locations} mentioned in the caption are noticeably absent from the image itself. This discrepancy indicates that the caption does not accurately represent the image, making it misleading or \textit{falsified news}.} & \\  

\bottomrule
\end{tabular}
\label{tab:discussion_case}
\end{table*}

To further demonstrate the effectiveness of \textbf{\texttt{EXCLAIM}}, we present a case study focusing on the challenge of place and entity verification in the context of OOC misinformation detection, as shown in Table~\ref{tab:discussion_case}.

In this case, the caption describes people cheering and taking pictures as the pope arrives to meet with representatives of the World of Work organization in Ciudad Juarez. The ground truth label (\textcolor{cyan}{GT}) indicates that this caption is falsified, meaning it does not match the image. While GPT-4o acknowledges the general alignment between the caption and the image, it ultimately states that the lack of clear identifying features or direct links to the specific event makes it difficult to confirm the accuracy of the caption. GPT-4o's response, though accurate in identifying uncertainty, remains superficial and lacks the capability to provide a decisive conclusion based on contextual evidence.

In contrast, \textbf{\texttt{EXCLAIM}} delivers a more nuanced analysis. The system observes the image of a crowd taking pictures or videos and identifies a group of individuals walking down a hallway, but no specific identifying features are visible. By retrieving and cross-referencing news captions, \textbf{\texttt{EXCLAIM}} confirms the geographic context of the event, identifying Ciudad Juarez as the location of the meeting. However, \textbf{\texttt{EXCLAIM}} also detects that the individuals and the event described in the caption are notably absent from the image itself. This discrepancy leads \textbf{\texttt{EXCLAIM}} to conclude that the caption does not align with the image, labeling the content as falsified. Unlike GPT-4o, which remains uncertain, \textbf{\texttt{EXCLAIM}} uses detailed verification mechanisms to identify the falsification.
Additional examples and further analysis are provided in the Appendix~\ref{app:case studies}, showcasing \textbf{\texttt{EXCLAIM}}'s performance across a variety of real-world contexts.
\section{Conclusion}

In this paper, we presented \textbf{\texttt{EXCLAIM}}, a novel framework that combines multi-granularity retrieval with a multi-agent reasoning system to address out-of-context misinformation. Through our self-constructed database and specialized agent collaboration, \textbf{\texttt{EXCLAIM}} demonstrates superior performance, achieving a 4.3\% accuracy improvement on the NewsCLIPpings benchmark.
 The framework's capability to analyze multi-modal inconsistencies at both the entity and event levels offers a more nuanced and robust approach to misinformation detection compared to existing methods. Looking ahead, future work could further enhance \textbf{\texttt{EXCLAIM}} by incorporating external knowledge bases and expanding its applicability to a broader range of misinformation detection challenges. Given its modular architecture, the framework holds significant potential to evolve into a comprehensive and scalable solution for multi-modal misinformation detection.

\newpage %
\section{Limitations}

\paragraph{Latency in Multi-Agent Collaboration:} The multi-agent reasoning architecture, while effective for explainability and systematic analysis, introduces additional computational overhead. This could limit the deployment of \textbf{\texttt{EXCLAIM}} in real-time applications where rapid decision-making is critical.

\paragraph{Challenges in Fine-Grained Visual Reasoning:} Despite leveraging advanced visual-textual alignment mechanisms, \textbf{\texttt{EXCLAIM}} occasionally struggles with fine-grained visual inconsistencies, particularly in tasks involving nuanced scene or person-level mismatches.


\bibliographystyle{acl_natbib}
\bibliography{custom}

\appendix


\section{Appendix}
\label{sec:appendix}

\subsection{Why not Open-source Model?}

\begin{table*}[t!]
\centering
\caption{Accuracy comparison (\%) between the GPT-4o and LLava Models.}
\label{table: discussion_llava}
\begin{tabular}{lcccc}
\toprule
\textbf{} & \textbf{GPT-4o-Latest} & \textbf{GPT-4o-mini} & \textbf{LLaVA-13B} & \textbf{LLaVA-7B}\\
\hline
\textbf{Accuracy}           & \cellcolor{lightgray}91.7\%         & 84.6\%          & 56.2\%     &   43.8\%  \\
\hline
\end{tabular}
\end{table*}

\begin{figure*}[t!]
    \centering
    \includegraphics[width=\linewidth]{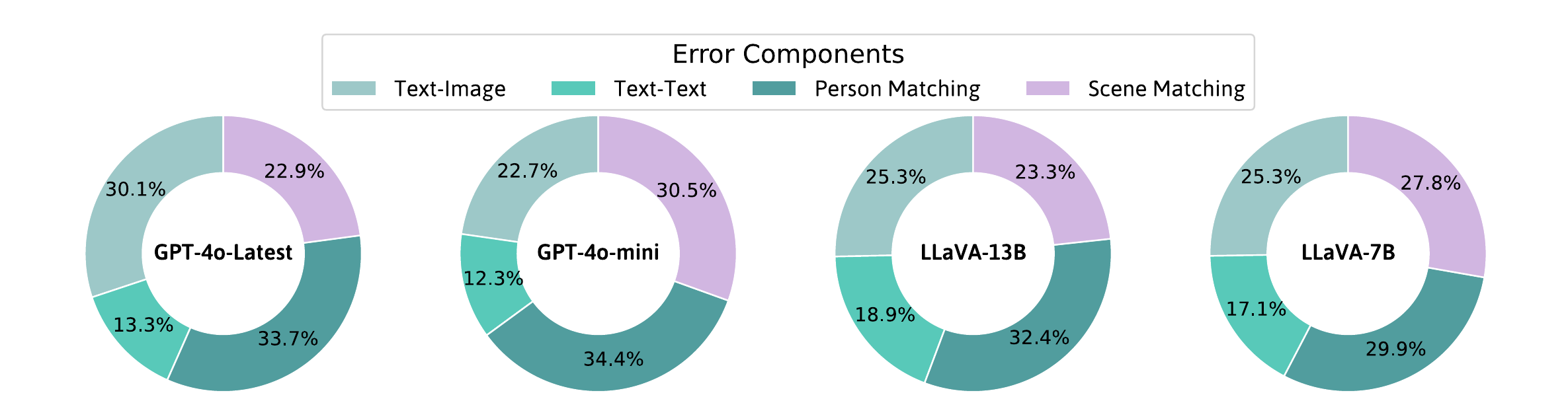}
    \caption{Error Distribution of GPT-4o and LLaVA Models on Different Type OOC Misinformation.}
    \label{fig:disucssion_errors}
\end{figure*}

In this section, we discuss the impact of replacing the base model in our multi-agent system with open-source alternatives. To better understand the implications of such a change, we conducted a detailed analysis using the four data types provided by the NewsCLIPpings dataset.  The NewsCLIPpings dataset defines four primary types of mismatches as described in the Section 4.1.1. Semantics Matching involves pairing images with captions that align in general content but differ in specific entities or events. This is split into two subtypes: \textbf{\textit{Text-Image}}, which retrieves images based on overall visual-textual similarity, and \textbf{\textit{Text-Text}}, where a semantically similar caption is first found, and the image from that caption is then mismatched with the original text. \textbf{\textit{Person Matching}} focuses on cases where the correct individual is depicted, but the person is placed in a misleading or unrelated context. Finally, \textbf{\textit{Scene Matching}} mislabels the broader setting or event, ensuring the environment looks similar but describes a different situation, excluding any references to individuals. For our evaluation, we maintained an equal distribution of 1,000 samples, with 250 examples from each category, to ensure a balanced and comprehensive assessment of model performance across these different misinformation scenarios.

Table~\ref{table: discussion_llava} shows a clear performance gap between open-source models like LLaVA 1.5~\citep{liu2023llava} and closed-source counterparts. Despite using the CLIP-ViT-L-336px architecture, LLaVA-7B and LLaVA-13B struggled with \textit{Person Matching} and \textit{Scene Matching} tasks, tasks requiring precise visual-textual alignment. Their smaller parameter sizes (7B and 13B) and shorter context windows limited their ability to process complex scenes. Prompt engineering yielded minimal improvements, emphasizing the architectural constraints in handling advanced multi-modal reasoning.

In contrast, closed-source GPT-4o models excelled across all OOC misinformation categories, as shown in Figure~\ref{fig:disucssion_errors}. Their larger parameter sizes and extended context windows allowed for better handling of intricate cross-modal relationships, especially in \textit{Scene Matching}, which requires deep contextual understanding. Additionally, the ease of deployment and regular updates of commercial models offer further advantages. Using state-of-the-art closed-source models improves the robustness of our misinformation detection system while avoiding the complexities of local deployment. Continuous updates ensure that our \textbf{\texttt{EXCLAIM}} framework remains at the forefront of multi-modal misinformation detection.

\subsection{Explainability Analysis}
\begin{table*}[h!]
\centering
\caption{ Average Rankings of Four Base Models for Logic and Explanation (Human and GPT-4o Evaluations). The best results for each test data are highlighted in bold.}
\label{table: discussion expla}
\begin{tabular}{lcccc}
\toprule
  \multirow{2}{*}{\textbf{Method}} & \multicolumn{2}{c}{\textbf{Human}} & \multicolumn{2}{c}{\textbf{GPT-4o}} \\ 
 \cmidrule{2-5} & Logic & Explanation & Logic & Explanation \\ \midrule
  LLaVA-7b & 3.60 & 3.01 & 3.05 & 2.85 \\
      LLaVA-13b & 3.20 & 3.45 & 3.55 & 3.50 \\
    GPT-4o-mini & 1.90  & 2.00  & 2.00  & 2.12 \\ 
    GPT-4o-Latest & \cellcolor{lightgray}\textbf{1.28} & \cellcolor{lightgray}\textbf{1.45} & \cellcolor{lightgray}\textbf{1.38} & \cellcolor{lightgray}\textbf{1.48} \\ \hline
\end{tabular}
\end{table*}
To assess the quality of explanations generated by the \textbf{\texttt{EXCLAIM}} framework, we conducted evaluations using both human evaluation and GPT-4o evaluation. For each of the 40 randomly selected test samples, both human evaluators and GPT-4o ranked the explanations generated by the four base models according to two criteria: \textbf{logical consistency (Logic)} and \textbf{explanatory quality (Explanation)}. Each model was assigned a rank from 1 (best) to 4 (worst) for each test case, and the average ranking across all samples was calculated for both logic and explanation.

The human evaluations were conducted by five undergraduate students from a science and engineering university program (three male and two female). These evaluators were recruited specifically for this study. They were provided with detailed guidelines and examples to ensure consistency in the evaluation process. Their academic background in STEM fields ensured they had sufficient analytical skills to assess logical consistency and explanatory quality effectively. All evaluators worked independently to minimize bias.

As shown in Table~\ref{table: discussion expla}, GPT-4o-Latest consistently achieved the best performance, with the lowest average rankings of 1.38 for logic and 1.48 for explanation in the GPT-4o evaluation. Human evaluators provided similar results, with average rankings of 1.28 for logic and 1.45 for explanation, further confirming the model’s strong reasoning capabilities and clarity. GPT-4o-mini, while slightly behind, still performed well, demonstrating the robustness of the GPT-4o architecture even in smaller-scale versions.
In contrast, LLaVA-13B and LLaVA-7B performed significantly worse, with higher average rankings across both criteria. LLaVA-13B had average rankings of 3.55 for logic and 3.50 for explanation in the GPT-4o evaluation, indicating difficulties in generating coherent reasoning. LLaVA-7B also struggled, with average rankings of 3.05 for logic and 2.85 for explanation.

These results highlight the superiority of GPT-4o models in producing explanations that are both logically sound and explainable, making them more suitable for complex multi-modal reasoning tasks, such as misinformation detection.

\subsection{Error Analysis Across Different Misinformation Categories}
\label{ssse:category_ana}
\begin{table*}[t!]
\centering
\caption{Distribution of Error Cases Across Different Categories in NewsCLIPpings Test Dataset.}
\begin{tabular}{lccc}
\toprule
\textbf{Category} & \textbf{Error Count} & \textbf{Error Rate (\%)} & \textbf{Primary Error Patterns} \\
\midrule
Text-Image & 177 & 33.40\% & Semantic similarity confusion \\
Person-Matching & 174 & 32.83\% & Contextual misalignment \\
Scene-Matching & 106 & 20.00\% & Environmental ambiguity \\
Text-Text & 73 & 13.77\% & Narrative consistency issues \\
\hline
Total & 530 & 100.00\% & - \\
\hline
\end{tabular}
\label{tab:error_distribution}
\end{table*}
To provide a more comprehensive understanding of \textbf{\texttt{EXCLAIM}}'s performance characteristics, we conducted a detailed analysis of error cases across different categories in the NewsCLIPpings dataset. Table \ref{tab:error_distribution} presents the distribution of errors across the four primary categories: \textbf{\textit{Text-Image}}, \textbf{\textit{Text-Text}}, \textbf{\textit{Scene-Matching}}, and \textbf{\textit{Person-Matching}}.

Our analysis reveals several noteworthy patterns in \textbf{\texttt{EXCLAIM}}'s error distribution. Text-Image mismatches constitute the largest proportion of errors (33.40\%), suggesting that the framework faces the greatest challenges in cases where semantic similarities between images and text are subtly misaligned. This is closely followed by Person-Matching errors (32.83\%), indicating that distinguishing individuals in different contexts remains a significant challenge despite our multi-agent approach.

Scene-Matching errors account for 20.00\% of the total errors, primarily occurring in cases where environmental elements share visual similarities but represent different events or contexts. The lowest error rate was observed in Text-Text matching (13.77\%), suggesting that \textbf{\texttt{EXCLAIM}} performs relatively well in detecting inconsistencies when dealing with purely textual semantic relationships.

These findings suggest potential areas for future improvement: 1) \textbf{Enhanced semantic reasoning capabilities:} Improving the system's ability to detect subtle semantic misalignments between images and text, particularly in cases where surface-level similarities mask contextual inconsistencies; 2) \textbf{Refined person-context association:} Strengthening the framework's capability to accurately track and verify person-specific contextual information across different temporal and spatial settings. 3)\textbf{Advanced scene understanding:} Developing more sophisticated mechanisms for distinguishing between visually similar but contextually different environments and events; 4)\textbf{Improved narrative consistency checking:} Enhancing the system's ability to verify and validate textual narrative consistency across different sources and contexts.

This error distribution analysis provides valuable insights for future iterations of the \textbf{\texttt{EXCLAIM}} framework and highlights specific areas where additional attention could yield significant improvements in overall system performance.

\subsection{Case Studies}
\label{app:case studies}

Table~\ref{tab:cases} demonstrates \textbf{\texttt{EXCLAIM}}'s verification capabilities through three representative examples. Each case highlights a distinct aspect of verification: character identification in sports event and temporal alignment verification. These examples illustrate how \textbf{\texttt{EXCLAIM}} conducts comprehensive analysis by leveraging multiple information dimensions beyond simple visual-textual matching.

\renewcommand\arraystretch{1}
\begin{table*}[t]
\centering
\small
\caption{Case studies comparing GPT-4o and \textbf{\texttt{EXCLAIM}}'s verification capabilities across character and temporal dimensions. Ground truth labels (\textcolor{cyan}{GT}) are provided.}
\begin{tabular}{lll}
\toprule
\multicolumn{2}{p{26pc}}{{\bf Caption}:Cleveland Cavaliers forward LeBron James dunks the ball during against the Atlanta Hawks in Game 4 of the Eastern Conference Finals. \color{cyan}{[GT: Pristine]}} & \makecell[c]{\multirowcell{14}{\includegraphics[width = 8pc]{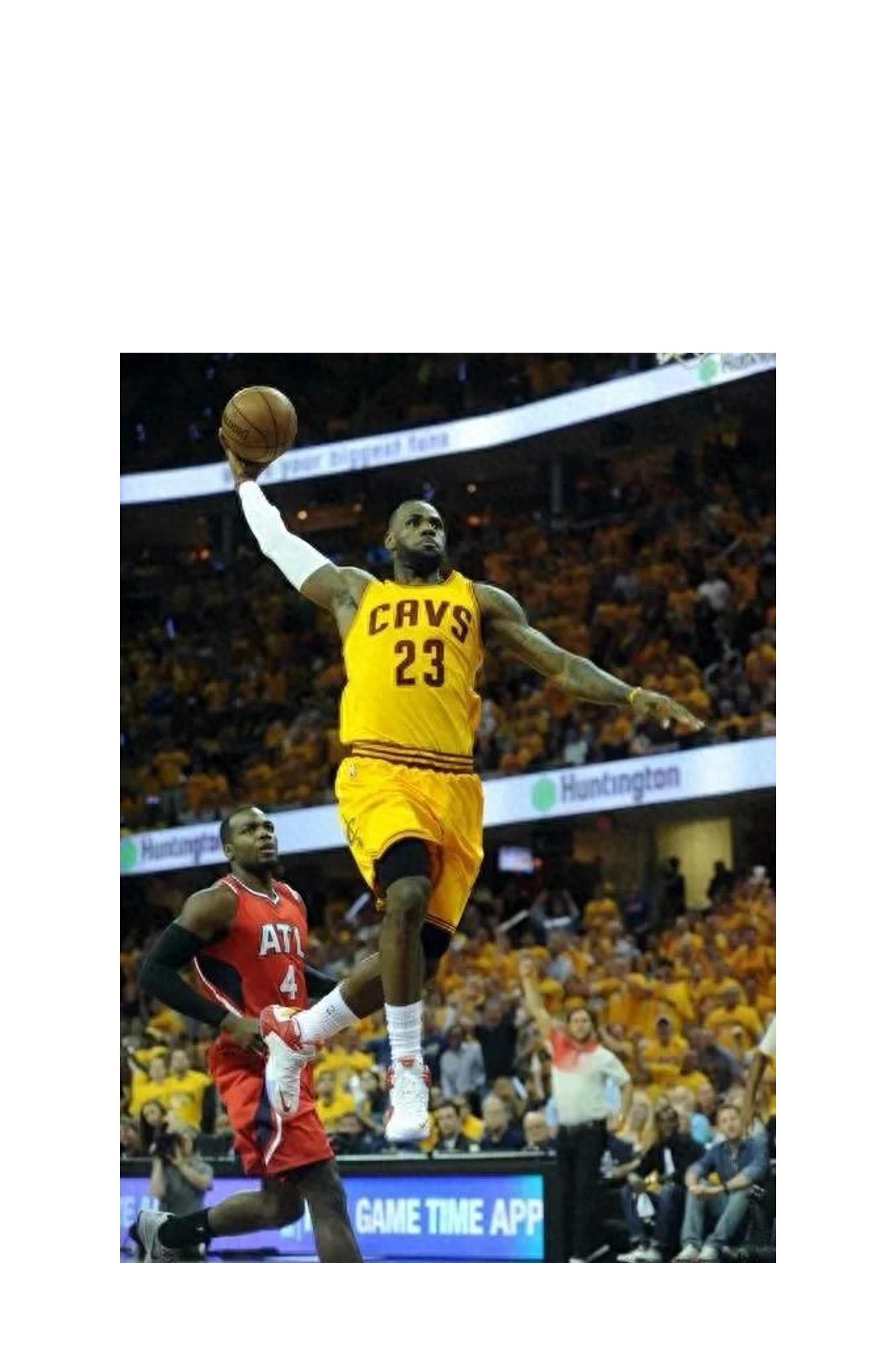}}}  \\
\cmidrule{1-2}
\multicolumn{1}{p{13pc}|}{{\bf GPT-4o}:  Yes, the caption matches the image. It shows a player wearing a Cleveland Cavaliers jersey with the number 23, who is dunking the basketball, and there appears to be an Atlanta Hawks player in the background. The description aligns with the visual context of the image. }  
&  
\multicolumn{1}{p{13pc}}{{\textbf{\texttt{EXCLAIM}}: In the given picture, a basketball player wearing a Cleveland Cavaliers jersey number 23 and another player from the opposing team (Atlanta Hawks jersey number 4) are visible. Based on the retrieved images, we can identify one of the players as \textbf{Jersey \#23 LeBron James}, a former Cleveland Cavaliers forward who played against the Atlanta Hawks in the NBA's 2015 Eastern Conference Finals. After comprehensive analysis, I think this is \textit{real} news.}} & 
\\

\midrule
\multicolumn{2}{p{26pc}}{{\bf Caption}: Florida Sen Marco Rubio speaks during a campaign rally in Minneapolis on Feb 23 2016. \color{cyan}{[GT: Pristine]}} & \makecell[c]{\multirowcell{14}{\includegraphics[width=8pc]{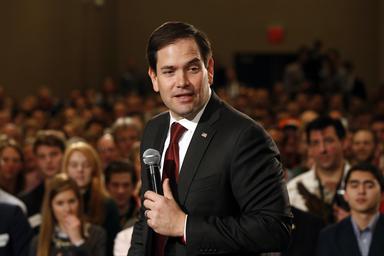}}}  \\
\cmidrule{1-2}
\multicolumn{1}{p{13pc}|}{{\bf GPT-4o}: Yes, the caption appears to match the image. It shows a man holding a microphone and speaking in front of an audience, which aligns with the description of Florida Sen. Marco Rubio speaking during a campaign rally. However, to confirm if the specific date and location (Feb 23, 2016, Minneapolis) are accurate, additional context would be required.}  
&  
\multicolumn{1}{p{13pc}}{{\textbf{\texttt{EXCLAIM}}}: In the given image, a man is speaking with a microphone in front of an audience. After comparing the retrieved image, we can confidently identify the individual as Marco Rubio. Following an event-level search, it has been confirmed that Marco Rubio participated in several presentations \textbf{during 2016}. Based on this comprehensive analysis, we conclude that this news is \textit{real}.} & \\

\bottomrule
\end{tabular}
\label{tab:cases}
\end{table*}

\subsection{Explanation Generation Capability Comparison}
\begin{table}[t!]
\centering
\caption{Explanation Generation Capability Comparison between Baselines and \textbf{\texttt{EXCLAIM}}.}
\label{table: explanation comparison}
\begin{tabular}{lc}
\toprule
\textbf{Method} & \textbf{Explanation Generation} \\
\midrule
EANN            & \textcolor{red}{\ding{55}}        \\
VisualBERT      & \textcolor{red}{\ding{55}}          \\
SAFE            & \textcolor{red}{\ding{55}}       \\
CLIP            & \textcolor{red}{\ding{55}}         \\
CCN             & \textcolor{red}{\ding{55}}           \\
DT-Transformer  & \textcolor{red}{\ding{55}}      \\
Neu-Sym detector & \textcolor{red}{\ding{55}}            \\
GPT-4o (zero-shot) &    \textcolor{green}{\ding{51}}       \\
GPT-4o (few-shot) &      \textcolor{green}{\ding{51}}   \\
SNIFFER  & \textcolor{green}{\ding{51}} \\
\hline
\textbf{\texttt{EXCLAIM}} \textbf{(ours)} & \textcolor{green}{\ding{51}}   \\
\hline
\end{tabular}
\end{table}

As shown in Table \ref{table: explanation comparison}, we compare the explanation generation capabilities of existing baselines and our proposed \texttt{EXCLAIM}. The results reveal a significant gap between most baseline methods and modern advanced models in their ability to generate explanations. Specifically, the majority of baseline methods (e.g., EANN, VisualBERT, SAFE) are marked with \textcolor{red}{\ding{55}}, indicating that they \textit{cannot generate explanations}. These models are primarily designed for misinformation detection tasks, with architectures that lack the capacity to produce explainable outputs. Even with potential modifications, enabling explanation generation in these models would require substantial architectural changes.

In contrast, models such as GPT-4o (in both zero-shot and few-shot settings), SNIFFER, and our proposed \texttt{EXCLAIM} are capable of generating explanations (\textcolor{green}{\ding{51}}). These models not only detect misinformation but also provide detailed justifications for their conclusions. Notably, \texttt{EXCLAIM} leverages multi-modal alignment mechanisms and knowledge-enhanced databases to produce high-quality explanations, significantly improving transparency and user trust in automated detection systems.

It is important to clarify the distinction between \textit{cannot generate explanations} and \textit{don’t generate explanations}. The former refers to models like EANN and VisualBERT, which inherently lack the architectural design to support explanation generation. The latter refers to scenarios where models, such as GPT-4o, may theoretically have the capability to generate explanations but are not explicitly configured to do so in certain tasks or settings.

The rapid advancements in multi-modal large language models (MLLMs) have been pivotal in enabling more powerful and explainable misinformation detection systems. By integrating both visual and textual modalities, these models excel at uncovering fine-grained inconsistencies and contextual misalignments, which are crucial for detecting out-of-context misinformation. Furthermore, their ability to provide detailed, explainable explanations not only improves detection accuracy but also enhances the transparency and reliability of the entire system.

In summary, our results highlight the transformative potential of modern AI models, particularly MLLMs, in bridging the gap between misinformation detection and explanation generation. Future research should focus on incorporating explanation capabilities into existing detection methods to build more robust and trustworthy systems.

\subsection{Comparison with SNIFFER Model}

In this section, we provide a detailed comparison between our proposed \textbf{\texttt{EXCLAIM}} framework and the SNIFFER model \cite{qi2024sniffer}, a prominent approach in the field of OOC misinformation detection. Both models leverage the power of MLLMs to tackle the challenges of OOC misinformation, yet they differ significantly in methodology, performance, explainability, and adaptability to various datasets, leading to distinct advantages and limitations.

From a methodological perspective, SNIFFER employs a two-stage instruction tuning approach, adapted from InstructBLIP, to refine its ability to align generic objects with news-domain entities and subsequently fine-tune its discriminatory powers for OOC misinformation detection. This process involves the integration of external knowledge through retrieval mechanisms, enabling SNIFFER to perform both internal checks (image-text consistency) and external checks (claim-evidence relevance), with the final decision produced through composed reasoning. While this is an effective approach, it introduces a reliance on external retrieval systems, which can introduce noise and latency in real-time applications. In contrast, \textbf{\texttt{EXCLAIM}} adopts a multi-agent architecture that decomposes the complex reasoning task into specialized subtasks, handled by agents responsible for retrieval, detection, and analysis. This modular structure not only enhances the interpretability of the system but also allows for more fine-grained verification through multi-granularity retrieval of both entity- and event-level information. By structuring its framework around a self-constructed multi-granularity database, \textbf{\texttt{EXCLAIM}} reduces dependency on external sources, offering a more efficient and unified approach to misinformation detection.

\paragraph{Performance}In terms of performance, both models demonstrate state-of-the-art capabilities, but \textbf{\texttt{EXCLAIM}} consistently outperforms SNIFFER across several benchmarks. SNIFFER reports an accuracy of 88.4\% on the NewsCLIPpings dataset, leveraging its external retrieval mechanisms to detect inconsistencies in OOC samples. However, \textbf{\texttt{EXCLAIM}} achieves an accuracy of 92.7\%, a significant improvement attributed to its multi-agent collaboration and multi-granularity retrieval system. This structured approach allows \textbf{\texttt{EXCLAIM}} to handle more subtle and complex OOC cases by cross-validating information across different granularities, thus providing a more robust detection mechanism. While SNIFFER’s retrieval-based methodology strengthens its performance, particularly in cases where external evidence is readily available, \textbf{\texttt{EXCLAIM}}’s internal verification process ensures that it remains highly effective even in scenarios where such evidence may be limited or noisy.

\paragraph{Explainability} Explainability is another critical dimension where the two models diverge. SNIFFER integrates its internal and external verification results to generate explanations, often relying on external evidence to justify its decisions. By incorporating web-based evidence, SNIFFER can provide detailed explanations that highlight the inconsistencies between the image and the text, such as misidentified entities or mismatched events. However, this reliance on external data can sometimes lead to overfitting to retrieved evidence, potentially complicating the interpretability of the decision-making process. \textbf{\texttt{EXCLAIM}}, on the other hand, enhances explainability through its multi-agent architecture, where each agent contributes specialized reasoning to the final output. The Retrieval Agent, Detective Agent, and Analyst Agent collaborate to ensure that the reasoning process is transparent and explainable at every stage. By ensuring that the decision-making process is broken down into distinct phases, \textbf{\texttt{EXCLAIM}} not only provides accurate judgments but also offers more structured and comprehensible explanations, further strengthened by the integration of multi-granularity data, which adds depth to its contextual understanding.

\paragraph{Adaptability} When considering the adaptability of these models to diverse datasets, \textbf{\texttt{EXCLAIM}}’s design offers a clear advantage. SNIFFER demonstrates strong generalization capabilities, as evidenced by its success across datasets such as News400 and TamperedNews, where it outperforms several baselines. However, its reliance on external retrieval introduces potential vulnerabilities to noisy or incomplete data, which can affect its overall robustness. \textbf{\texttt{EXCLAIM}}’s multi-granularity database construction and internal verification process allow it to adapt more effectively to different types of misinformation across various contexts. By cross-referencing data at both the entity and event levels, \textbf{\texttt{EXCLAIM}} ensures that it can consistently maintain high performance across diverse datasets without being overly dependent on the availability of external evidence. This adaptability makes \textbf{\texttt{EXCLAIM}} particularly well-suited for real-world applications where external sources may not always provide reliable or timely information.

\paragraph{Efficiency} Finally, with respect to efficiency, \textbf{\texttt{EXCLAIM}}’s multi-agent system provides a significant advantage. SNIFFER’s reliance on external tools and web-based retrieval can introduce latency, particularly in real-time or large-scale applications where the availability and quality of external data are critical. In contrast, \textbf{\texttt{EXCLAIM}}’s internal multi-agent collaboration and self-constructed database allow it to operate more efficiently. The modular design of \textbf{\texttt{EXCLAIM}}’s agents ensures that each step of the verification process is optimized for speed and accuracy, making it more suitable for real-time OOC misinformation detection. By reducing dependency on external retrieval, \textbf{\texttt{EXCLAIM}} minimizes computational overhead while maintaining high detection accuracy, a crucial factor for practical deployment in fast-paced information environments.

In conclusion, while both SNIFFER and \textbf{\texttt{EXCLAIM}} represent significant advancements in the detection of OOC misinformation, \textbf{\texttt{EXCLAIM}}’s innovative multi-agent architecture, multi-granularity retrieval system, and focus on internal verification offer superior performance, interpretability, and adaptability. These differences highlight \textbf{\texttt{EXCLAIM}}’s robustness in handling complex misinformation scenarios and its potential for real-world application, setting it apart as a more comprehensive and efficient solution for OOC misinformation detection.

\end{document}